\documentclass[journal=jctcce,manuscript=article]{achemso}

\usepackage{amsmath}
\usepackage{amsfonts}
\usepackage{mathtools}
\usepackage{cancel}
\usepackage{chemformula} 
\usepackage[T1]{fontenc} 
\usepackage{soul}

\author{J. L. Alonso}
\affiliation{Departamento de F\'{\i}sica Te\'orica, Facultad de Ciencias, Universidad de Zaragoza,  Campus San Francisco, 50009 Zaragoza (Spain)}
\alsoaffiliation{Instituto de Biocomputaci{\'{o}}n y F{\'{\i}}sica de Sistemas Complejos (BIFI), 
Universidad de Zaragoza,  Edificio I+D, Mariano Esquillor s/n, 50018 Zaragoza (Spain)}
\alsoaffiliation{Centro de Astropart{\'{\i}}culas y F{\'{\i}}sica de Altas Energ{\'{\i}}as, Facultad de Ciencias, Universidad de Zaragoza,  Campus San Francisco, 50009 Zaragoza (Spain)}

\author{C. Bouthelier}
\email{cbouthelier@unizar.es}
\affiliation{Departamento de F\'{\i}sica Te\'orica, Facultad de Ciencias, Universidad de Zaragoza,  Campus San Francisco, 50009 Zaragoza (Spain)}
\alsoaffiliation{Instituto de Biocomputaci{\'{o}}n y F{\'{\i}}sica de Sistemas Complejos (BIFI), 
Universidad de Zaragoza,  Edificio I+D, Mariano Esquillor s/n, 50018 Zaragoza (Spain)}
\alsoaffiliation{Centro de Astropart{\'{\i}}culas y F{\'{\i}}sica de Altas Energ{\'{\i}}as, Facultad de Ciencias, Universidad de Zaragoza,  Campus San Francisco, 50009 Zaragoza (Spain)}

\author{A. Castro}
\affiliation{Fundación ARAID, Zaragoza (Spain)}
\alsoaffiliation{Instituto de Biocomputaci{\'{o}}n y F{\'{\i}}sica de Sistemas Complejos (BIFI), 
Universidad de Zaragoza,  Edificio I+D, Mariano Esquillor s/n, 50018 Zaragoza (Spain)}
\alsoaffiliation{Departamento de F\'{\i}sica Te\'orica, Facultad de Ciencias, Universidad de Zaragoza,  Campus San Francisco, 50009 Zaragoza (Spain)}

\author{J. Clemente-Gallardo}
\affiliation{Departamento de F\'{\i}sica Te\'orica, Facultad de Ciencias, Universidad de Zaragoza,  Campus San Francisco, 50009 Zaragoza (Spain)}
\alsoaffiliation{Instituto de Biocomputaci{\'{o}}n y F{\'{\i}}sica de Sistemas Complejos (BIFI), 
Universidad de Zaragoza,  Edificio I+D, Mariano Esquillor s/n, 50018 Zaragoza (Spain)}
\alsoaffiliation{Centro de Astropart{\'{\i}}culas y F{\'{\i}}sica de Altas Energ{\'{\i}}as, Facultad de Ciencias, Universidad de Zaragoza,  Campus San Francisco, 50009 Zaragoza (Spain)}

\author{J. A. Jover-Galtier}
\affiliation{Centro Universitario de la Defensa de Zaragoza, Academia General Militar, 50XXX Zaragoza (Spain)}
\alsoaffiliation{Instituto de Biocomputaci{\'{o}}n y F{\'{\i}}sica de Sistemas Complejos (BIFI), 
Universidad de Zaragoza,  Edificio I+D, Mariano Esquillor s/n, 50018 Zaragoza (Spain)}

\title[]
{About the computation of finite temperature ensemble averages of hybrid quantum-classical systems
with Molecular Dynamics}

\abbreviations{IR,NMR,UV}

\begin{document}







\begin{abstract}
Molecular or condensed matter systems are often well approximated by
hybrid quantum-classical models: the electrons retain their quantum
character, whereas the ions are considered to be classical particles.
We discuss various alternative approaches for the computation of
equilibrium (canonical) ensemble averages for observables of these
hybrid quantum-classical systems through the use of molecular dynamics
(MD) -- i.e. by performing dynamics in the presence of a thermostat
and computing time-averages over the trajectories.  Often, in
classical or ab initio MD, the temperature of the electrons is ignored
and they are assumed to remain at the instantaneous ground state given
by each ionic configuration during the evolution. Here, however, we
discuss the general case that considers both classical and quantum
subsystems at finite temperature canonical equilibrium.  Inspired by
a recent formal derivation for the canonical ensemble for quantum
classical hybrids, we discuss previous approaches found in the
literature, and provide some new formulas.
\end{abstract}

\section{\label{sec:introduction} Introduction}

Molecular Dynamics (MD)~\cite{Rapaport2004} is conventionally
considered to be the theoretical description of molecular or condensed
matter systems that assumes the nuclei to be classical
particles. Therefore, these move according to Newton's equations, in
the presence of their mutual interaction and of a force that somehow
approximates the electron influence. In its traditional formulation
(\emph{classical} MD), the forces are parameterised in some analytical
expressions that have been carefully developed over the years, and the
numerical problem amounts to the propagation of a purely classical
Hamiltonian system with a predefined potential function. The so-called
\emph{ab initio} or \emph{first-principles} MD~\cite{Marx2009}
substitutes those analytical force definitions by the \emph{on the
  fly} calculation of the quantum electronic structure problem, that
provides the forces on the ions due to the electrons in a more precise
-- yet more costly -- manner. Still, a first principles MD simulation
also consists in the integration of a purely classical problem, even
though one needs to use quantum mechanics to obtain the forces at each
time step. In both classical and first principles MD, the electrons
are usually assumed to remain in their ground state,
\emph{adiabatically} adapting to the ions as they move. Therefore,
both the classical MD and the (Born-Oppenheimer ground-state) first-principles MD are
not, strictly speaking, hybrid quantum-classical dynamics. These methods have been
widely employed for both equilibrium and out-of-equilibrium problems.

In the equilibrium case, when studying a system in, e.g., the canonical
ensemble, one is normally not interested in the particular trajectory
followed by a microstate, but on the ensemble average values of a
given property. The MD simulations are then used to compute the
multi-dimensional integrals that define those averages, by
substituting them with time-averages over dynamical trajectories of
the system -- typically, coupled to a
\emph{thermostat}~\cite{Tuckerman2010}.  But in any case, despite the
finite temperature, normally the electrons are assumed to be frozen in
the ground state.  This may be a very good approximation if the
electronic excited state energies are far higher than the thermal
energies at that temperature. Yet in many circumstances those excited
states cannot be ignored.

Out of equilibrium, this may in fact happen very frequently. For
example, when dealing with photo-chemistry, that naturally involves
electronic excitations.  This situation calls for a
\emph{non-adiabatic} extension of the previous MD concept, that allows
for ``live'' electrons: the dynamics must be that of a true hybrid
quantum-classical model, in which both classical and quantum particles
evolve simultaneously through a set of coupled equations. Two
prototypical examples of truly hybrid dynamics are Ehrenfest equations~\cite{Bornemann1996}
and surface hopping~\cite{Tully1998} (in this latter case, the
electronic motion is stochastic and consists of
``jumps'' between the adiabatic eigenstates).

In equilibrium, one may also need to lift the approximation of ground
state electrons, if the temperature is high enough that the thermal
population of the excited states is not negligible. This is the
situation discussed in this article. In molecular physics this may
happen rarely, but in condensed matter, the metallic or near metallic
systems naturally call for a computation of ensemble averages that
acknowledges the non-zero population of electronic excited states at
non-zero temperature, even if low.  In any case, this situation begs
the questions: what are the canonical ensemble averages for
observables of hybrid systems, and can one compute them using MD? The
standard procedure used within classical or adiabatic first principles
MD is no longer directly applicable if one is simultaneously
propagating nuclei and electrons. The idea of assuming ergodicity and
attaching a thermostat to the dynamics, a concept designed for purely
classical systems, is dubious at best.

This issue has been addressed by performing MD propagating only the
classical nuclei, but somehow incorporating the electronic
temperature in the definition of the forces, instead of deriving them
by merely assuming the electrons to be in the ground state. One
possible route, based on the use of density-functional theory, was
theorised by Alavi \latin{et al.}~\cite{Alavi1994}. Their method was
based on the use of density-functional theory~\cite{Burke2012} (DFT)
to solve for the electronic structure problem. In particular, on the
finite temperature extension of DFT (FTDFT)~\cite{Mermin1965,PribramJones2014}, that substitutes the
ground-state energy functional by a free energy functional. Then, to
perform first principles MD at finite temperature, the forces used to
propagate the classical ions are given by the gradient of this free
energy functional.  In essence, this is the same idea that
underlies the approach given in Refs.~\citenum{Alonso2010, Alonso2012},
except that in this latter case the formulation is general, and not tied to
DFT. The fact that the electronic free energy -- considered at fixed
nuclear configurations -- can be viewed as an effective classical
Hamiltonian from which the hybrid quantum-classical partition function
can be computed was already found by Zwanzig~\cite{Zwanzig1957}. DFT
is in fact the most common electronic structure method for the purpose
of performing ab initio MD. The inclusion of electronic temperature
effects is therefore usually managed with some form of FTDFT. In
practice, this procedure consists of using a Fermi-Dirac distribution
for the population of the Kohn-Sham orbitals that constitute the
fictitious auxiliary non-interacting system employed to substitute the
true interacting many-electron problem. The resulting density is used
to compute the ionic forces, in lieu of the ground-state density. The
procedure should be completed with the use of temperature-dependent
exchange-and-correlation functionals, but this is often ignored, as the
development of these functionals has proved to be very difficult.

In this work, we discuss this and other possible routes to obtain the
rigorous canonical ensemble averages through the use of thermosthatted
MD.  The goal is to establish a clear theoretical link between the
definition of hybrid ensemble averages and the manners that one can
use to compute them using some form of MD with a thermostat. The basic
idea consists of generating an ensemble with some form of MD, even if
the generated ensemble is wrong, and then using a reweighting formula
to compute the right averages.  From the analysis, it emerges that, in
fact, various possibilities exist. The dynamics for the classical
particles moving on the free energy surface will be shown to be a very
particular case of this general class. The relative efficiency of the
various options may depend on the particular system and choice of
electronic structure method. The idea of performing wrong or
ficititious dynamics, and then correcting with some reweighting
procedure, has been used in the past in the field of MD mostly with the
objective of accelerating rare events -- see for example
Refs.~\citenum{Hamelberg2004,Oliveira2006}. In this work we extend 
the same idea to the simulation of hybrid quantum-classical systems.

In Section~\ref{sec:hce} we present the expressions for the ensemble
averages that constitute the target of the current
work. Section~\ref{sec:ehrenfest} discusses the possibility of
computing these using a hybrid quantum-classical non-adiabatic MD such
as Ehrenfest dynamics. Although the na{\"{\i}}ve computation of
time-averages over thermostatted Ehrenfest dynamics trajectories lead
to wrong results (as already noted in earlier
works~\cite{Mauri1993,Parandekar2005,Parandekar2006}), we discuss ways
to correct this issue. Section~\ref{sec:md} discusses approaches based
on classical-only MD propagations.

\section{\label{sec:hce} The canonical ensemble of hybrid quantum-classical systems}

We start by recalling which are the ensemble averages that we are
addressing in this work.

The first step should be to clarify the mathematical description of a
hybrid model, an issue that is not at all obvious, as demostrated by
the various proposals that have been put forward, and by the
discussions about their internal consistency~\cite{Prezhdo1997,
  Kisil2005, Prezhdo2006, Salcedo2007, Agostini2007, Kisil2010,
  Agostini2010, Hall2008, Buric2013b, Peres2001, Terno2006,
  Salcedo1996, Gil2017, Caro1999, Diosi2014, Elze2012,
  Aleksandrov1981, Kapral1999}.  We will however assume the following
very broad assumptions.  The classical part is described by a set of
position $Q \in \mathbb{R}^n$ and momentum $P \in \mathbb{R}^n$
variables, that we will hereafter collectively group as $\xi = (Q,
P)$. Normally, they correspond to $N$ particles, such that $n=3N$ in
three dimensions. The quantum part is described by a complex Hilbert
space $\mathcal{H}$. The observables of the full hybrid system are Hermitian operators on
$\mathcal{H}$ that may depend parametrically on the classical
variables, $\hat{A}(\xi) : \mathcal{H} \to \mathcal{H}$. Some
observables may refer only to the classical subsystem; in that case
they are just $\xi$-functions times the identity, i.e. $\hat{A}(\xi) =
A(\xi)\hat{I}$. If, on the contrary, they refer to the quantum
subsystem only, they are operators that lack the
$\xi$-dependence. In any other case, a hybrid observable couples the quantum and classical
parts to each other. The most important one is the Hamiltonian
$\hat{H}(\xi)$. Although its precise form is not important for the
following discussion, as an example we write here the typical
definition of this Hamiltonian for a set of $N_{\rm e}$ quantum
electrons and $N$ nuclei:
\begin{equation}
\label{eq:enham1}
\hat{H}(Q, P) = \left(\sum_{I=1}^N \frac{\vec{P}_I^2}{2M_I}\right)\hat{I} + \hat{H}_{\rm e}(Q)\,.
\end{equation}
The first term is the kinetic energy of the classical particles, whereas the
second part is
\begin{equation}
\label{eq:enham2}
\hat{H}_{\rm e}(Q) = \sum_{i=1}^{N_{\rm e}}\frac{\hat{\vec{p}}_i^2}{2m_e}
+ \hat{V}_{\rm en}(Q) + V_{\rm nn}(Q)\hat{I}\,,
\end{equation}
where the first term is the kinetic electronic operator, the second term is the electron-nucleus
interaction potential, and the last (purely classical) term is the nucleus-nucleus interaction potential.

Ensembles of hybrid quantum-classical systems can be described~\cite{Aleksandrov1981,Kapral1999,Alonso2020} by
$\xi$-dependent density matrices, $\hat{\rho}(\xi)$, normalized as:
\begin{equation}
\int{\rm d}\mu(\xi) {\rm Tr}\hat{\rho}(\xi) = 1\,.
\end{equation}
These fully
 characterize the ensemble, i.e. they permit to
obtain the probabilities associated to any measurement. For
example, the probability associated to finding the classical
subsystem at $\xi$, \emph{and} measuring $a$ for observable
$\hat{A}(\xi)$, is given by ${\rm
  Tr}\left[\hat{\rho}(\xi)\hat{\pi}_a(\xi)\right]$, where
$\hat{\pi}_a(\xi)$ is the projector associated to the eigenvalue $a$
of $\hat{A}(\xi)$. Or, the probability density associated to the
classical subsystem, regardless of the quantum part, is given by
$F_C(\xi) = {\rm Tr}\hat{\rho}(\xi)$. Likewise, given any observable $\hat{A}$, the
ensemble average is given by:
\begin{equation}
\langle \hat{A} \rangle_{\hat{\rho}} = \int{\rm d}\mu(\xi) {\rm Tr}\left[ \hat{A}(\xi)\hat{\rho}(\xi) \right]\,.
\end{equation}

One route to the definition of equilibrium ensembles is the principle
of maximization of entropy (MaxEnt). Recently, we argued~\cite{Alonso2020} that the proper
definition of entropy for a hybrid quantum-classical system must be:
\begin{equation}
S[\hat{\rho}] = - k_B \int {\rm d}\mu(\xi) {\rm Tr}\left[ \hat{\rho}(\xi)\log\hat{\rho}(\xi)\right]\,.
\end{equation}
Likewise, we also showed that the maximization of this entropy, subject to the constraint
of a given value for the average energy, leads to the \emph{hybrid
canonical ensemble}:
\begin{align}
\hat{\rho}_{\rm HC}(\xi) &= \frac{1}{Z_{\rm HC}(\beta)}e^{-\beta \hat{H}(\xi)}\,,
\\
Z_{\rm HC}(\beta) &= \int{\rm d}\mu(\xi) {\rm Tr} e^{-\beta\hat{H}(\xi)}\,.
\end{align}
Therefore, the canonical ensemble average of any observable is:
\begin{equation}
\label{eq:hybavg}
\langle \hat{A} \rangle_{\rm HC} (\beta) = \frac{1}{Z_{\rm HC}(\beta)} \int{\rm d}\mu(\xi) {\rm Tr}
\left[ \hat{A}(\xi)e^{-\beta\hat{H}(\xi)}  \right]\,.
\end{equation}
The computation of these averages is challenging. First, depending on
the model and quantum level of theory used, the calculation of the
traces, that in principle require all excited states, can be
problematic. But, more importantly, the integral over the classical
phase space is difficult because of its very large dimensionality
($6N$ for $N$ classical particles in 3D).

This latter problem is of course akin to the one encountered when
studying purely classical systems. It is therefore natural to ask
whether it is possible to circumvent it by doing some form of MD.

\section{\label{sec:ehrenfest} The failure of Ehrenfest dynamics, and some ways to correct it}

One possibility that immediately comes to mind is the use of a hybrid
quantum-classical MD, such as Ehrenfest's, and attaching a thermostat
in order to simulate the presence of a bath that would permit to
generate the canonical ensemble along the trajectory. In other words, replicating
the procedure invented for ``standard'' MD, but using a hybrid dynamics
that requires the explicit propagation of the electrons.

It was soon realized, however, that this procedure leads to wrong
ensemble averages~\cite{Mauri1993,Parandekar2005,Parandekar2006}.  In
the following, we will reexamine this fact in the light of the
Hamiltonian character of Ehrenfest dynamics. This analysis will help
to understand the correction procedures that in fact permit to
use this dynamics to obtain the true ensemble averages.

\subsection{\label{subsec:hamdynamics} Fast recap of Hamiltonian dynamics}

Let us first recap the basics of Hamiltonian theory. A system can be
characterized by providing a phase space $\mathcal{M}$ of even
dimension $2n$. {\color{black} The Poisson bracket is an operation defined over functions in
this phase space (the observables), which in
the canonical coordinates $(q,p)\in\mathcal{M}$ reads:
\begin{equation}
\label{eq:poisson}
\lbrace A,B\rbrace = \sum_{i=1}^n\left[
\frac{\partial A}{\partial q_i}\frac{\partial B}{\partial p_i} - 
\frac{\partial A}{\partial p_i}\frac{\partial B}{\partial q_i}
\right]\,.
\end{equation}

The dynamics is determined by the definition of a Hamiltonian function
$H$: the equations of motion for the coordinates $q_i$ or $p_i$ of
any state in $\mathcal{M}$ are $\dot{q}_i = \lbrace q_i, H\rbrace$
and $\dot{p}_i = \lbrace p_i, H\rbrace$, or 
equivalently, as they are more often encountered, in the form of Hamilton's equations:
\begin{align}
\dot{q}_i = \frac{\partial H}{\partial p_i}\,,
\\
\dot{p}_i = -\frac{\partial H}{\partial q_i}\,.
\end{align}

If there is no certainty about the system state, instead of a single
point, one must use a probability distribution $\rho(q,p,t)$ 
defined over the phase space, also known as an ensemble. This distribution
may change in time, according to Liouville's equation:
\begin{equation}
\frac{\partial \rho}{\partial t} = \lbrace H, \rho\rbrace\,.
\end{equation}

The entropy of any ensemble can be computed as (hereafter, we will group
all variables $q,p$ as $y$):
\begin{equation}
\label{eq:classentropy}
S[\rho] = -k_B \int{\rm d}\mu(y) \rho(y)\log\rho(y)\,.
\end{equation}
}
The maximization of this entropy over all possible ensembles subject
to the constraint of a given Hamiltonian ensemble average or energy,
$\langle H \rangle_\rho = \int{\rm d}\mu(y) H(y)\rho(y)$, leads
to the canonical ensemble~\cite{Reichl2016}:
\begin{align}
\label{eq:gibbs}
\rho_{\rm CC}(y) &= \frac{1}{Z_{\rm CC }(\beta)}e^{-\beta H(y)}\,,
\\
Z_{\rm CC}(\beta) &= \int{\rm d}\mu(y)\;e^{-\beta H(y)}\,.
\end{align}
Here, $\beta = \frac{1}{k_B T}$ is inversely proportional to the temperature $T$, ``CC'' stands
for ``classical canonical'', $Z_{\rm CC}(\beta)$ is the partition function, and the integrals
extend over all phase space. This is an equilibrium ensemble, lacking the time-dependence because
it is stationary: $\lbrace H, \rho_{\rm CC}\rbrace = 0$.

The averages, for any observable $A$, over this canonical ensemble are then given by:
\begin{equation}
\label{eq:classavg}
\langle A \rangle_{\rm CC}(\beta) = \frac{1}{Z_{\rm CC}(\beta)}\int{\rm d}\mu(y)\;e^{-\beta H(y)}A(y)\,,
\end{equation}
The obvious numerical difficulty of computing
these very high-dimensional integrals can then be circumvented by integrating a single dynamical trajectory, and using
the ergodic hypothesis to identify a time average with the phase space integral:
\begin{equation}
\label{eq:classergodic}
\langle A \rangle_{\rm CC}(\beta) = \lim_{t_f\to\infty}\frac{1}{t_f}\int_0^{t_f}\!\!{\rm d}t\; A(y^\beta(t))\,,
\end{equation}
where $y^\beta(t)$ is a trajectory obtained by solving the equations of the motion, modified
with a thermostat, i.e.:
\begin{equation}
\dot{y}^\beta_i = \lbrace y^\beta_i,H\rbrace + X^\beta_i(t)\,,
\end{equation}
Here, we have symbolically added to the Poisson bracket $\lbrace\cdot,\cdot\rbrace$ a thermostat $X_\beta(t)$ 
(it may represent Langevin's stochastic term~\cite{Lemons1997,Tuckerman2010}, a Nose-Hoover chain~\cite{Martyna1992}, etc.)

\subsection{\label{subsec:schehr} {\color{black}Schr{\"{o}}dinger  dynamics as  a Hamiltonian system}}

{\color{black} The theory summarized in subsection~\ref{subsec:hamdynamics} can be applied to any Hamiltonian
dynamics -- for example, to Schrödinger's equation, which despite its quantum character,
is a ``classical'' Hamiltonian system from a mathematical perspective. We summarize this
fact here -- for the mathematical conditions and functional spaces (both finite and infinite dimensional) on which this formalism
can be applied, see [\citenum{Chernoff1974,Kibble1979}]; in [\citenum{Heslot1985}], one can follow
the standard approach that is summarized here.}

{\color{black}Indeed, Schrödinger's equation ($\hbar = 1$ is assumed throughout this paper),}
\begin{equation}
\label{eq:schroedinger}
i\frac{\rm d}{{\rm d}t}\vert\psi(t)\rangle = \hat{H}\vert\psi(t)\rangle\,,
\end{equation}
{\color{black} is easy to rewrite as a set of Hamiltonian equations.} First,
one expands the wavefunction in an orthonormal basis $\lbrace \varphi_i\rbrace_i$, and rewrites
Schr{\"{o}}dinger's equation for the coefficients $c_i = \langle\varphi_i\vert\psi\rangle$:
{\color{black}
\begin{equation}
\dfrac{d}{dt} \begin{pmatrix}
c_1\\
c_2\\
\vdots \\
c_n
\end{pmatrix} = \begin{pmatrix} 
H_{11} & H_{12} & \cdots & H_{1n} \\
H_{21} & \ddots & \ddots & H_{2n} \\
\vdots & \ddots & \ddots & \vdots \\
H_{n1} & \cdots & \cdots & H_{nn}
\end{pmatrix} 
\begin{pmatrix}
c_1\\
c_2\\
\vdots \\
c_n
\end{pmatrix}\,,
\end{equation}}
where $H_{ij}$ are the elements of the Hamiltonian matrix $H_{ij} = \langle
\varphi_i \vert\hat{H}\vert\varphi_j\rangle$. We define
a set of ``position'' and ``momenta'' variables by taking
the real and imaginary parts of this coefficients, respectively:
$c = \frac{1}{\sqrt{2}}(q + ip)$. One may then show{\color{black}\cite{Kibble1979,Heslot1985}} that Eq.~(\ref{eq:schroedinger})
is equivalent to
\begin{align}
\label{qdynamics}
\dot{q}_i &= \frac{\partial  H}{\partial p_i}\,,
\\
\label{pdynamics}
\dot{p}_i &= -\frac{\partial  H}{\partial q_i}\,,
\end{align}
i.e. Hamiltonian's equations, defining $H(q,p)$ as the
expectation value of $\hat{H}$ for the wavefunction determined by the
$(q,p)$ coefficients: $H(q, p) = \langle\psi(q,p)\vert\hat H \vert\psi(q,p)\rangle$.
{\color{black} Note that these ``position'' and ``momemtum'' variables should not
be given any particular physical meaning, forcing any analogy with classical mechanics.}

The Poisson bracket can then be defined in the usual way
{\color{black} [Eq.~(\ref{eq:poisson})]. We will use the notation
$\lbrace\cdot,\cdot\rbrace_Q$ for this Poisson bracket defined
in this new ``quantum'' phase space, $\mathcal{M}_Q$, defined
by the variables $(q, p)$.}
The system dynamics then reduces to:
{\color{black} \begin{equation}
\label{QuantumHamiltonEquation}
\dot f = \lbrace f,  H\rbrace_Q\,,
\end{equation}}
{\color{black} for any function $f$ defined in $\mathcal{M}_Q$.} For the particular case of
the coordinate functions $q$ and $p$, one obtains the Hamilton
equations above. Taking into account the dependence of $\psi(q,p)$, they are
entirely equivalent to Schr{\"{o}}dinger's equation {\color{black}(see \citenum{Kibble1979,Heslot1985} for details).}

Gibbs canonical ensemble {\color{black}associated with the expectation value of the energy $H(q,p)$}, Eq.~(\ref{eq:gibbs}), is stationary under
the dynamics, and in principle one could attach one of the typical {\color{black}classical-type}
thermostats to Schr{\"{o}}dinger equation, and produce this ensemble
through a trajectory. If one did that, however, the resulting ensemble
averages would be wrong, because Gibbs ensemble is not the true
quantum canonical ensemble, which is defined through a density matrix as:
\begin{align}
\hat{\rho}_{\rm QC} &= \frac{1}{Z_{\rm QC}(\beta)}e^{-\beta\hat{H}}\,,
\\
Z_{\rm QC}(\beta) &= {\rm Tr} e^{-\beta\hat{H}}\,.
\end{align}
This is the density matrix that maximizes the von Neumann entropy,
\begin{equation}
S[\hat{\rho}] = -k_B {\rm Tr}(\hat{\rho}\log\hat{\rho})\,,
\end{equation}
which is the real entropy of a quantum system, and not
Eq.~(\ref{eq:classentropy}), from which the classical canonical
ensemble is derived. It is obvious that the fact that the
``thermostatted'' Schr{\"{o}}dinger dynamics does not produce the
correct thermal averages is not a defect of Schr{\"{o}}dinger
equation, but results of the erroneous application of a technique
invented for classical systems to quantum ones.

{\color{black}
\subsection{\label{subsec:ehr} Ehrenfest dynamics as a Hamiltonian system}}

{\color{black}
Ehrenfest dynamics, usually introduced as a partial classical limit of the full-quantum
dynamics~\cite{Bornemann1996} , constitutes also a Hamiltonian system,
as shown for example in [\citenum{Bornemann1996,Alonso2011,Alonso2018}]. We summarize here
this fact.
}

We consider a hybrid quantum-classical
system, as defined in Section~\ref{sec:hce}.  The classical variables
$\xi = (Q, P)$ define a classical phase space $\mathcal{M}_C$, whereas the
quantum variables $\eta=(q,p)$, associated to a wavefunction $\psi(\eta)$ as
explained above, define a quantum phase space $\mathcal{M}_Q$. We may
put these together and define a full, hybrid phase space:
\begin{equation}
\mathcal{M} = \mathcal{M}_C \times \mathcal{M}_Q\,.
\end{equation}

Furthermore, we may define a Poisson bracket for functions defined on this full
hybrid space by adding the two classical and quantum brackets, defined
over the $\xi$ and $\eta$ variables, respectively:
\begin{align}
\label{eq:hybpoisson}
\nonumber
\lbrace A, B \rbrace_H &= \lbrace A, B \rbrace_C + \lbrace A, B \rbrace_Q
\\
 &= {\color{black} \sum_i \left(\partial_{Q_i}A\partial_{P_i}B-\partial_{P_i}A\partial_{Q_i}B\right) + 
    \sum_i \left(\partial_{q_i}A\partial_{p_i}B-\partial_{p_i}A\partial_{q_i}B\right)\,. }
\end{align}
{\color{black}
Thus, the classical bracket derivates only with respect to the classical coordinates $(Q,P)$,
and the quantum bracket with respect to the quantum ones $(q,p)$.}
It is a well known result of {\color{black}Poisson geometry} that the
addition of two brackets {\color{black}in a Cartesian product} results in a bracket that fulfills all the
necessary properties.

Finally, given any hybrid observable $\hat{A}(\xi)$, we may define a real function
over $\mathcal{M}$ as:
\begin{equation}
A(\eta, \xi) = \langle \psi(\eta) \vert \hat{A}(\xi)\vert \psi(\eta) \rangle
\end{equation}
{\color{black}
If, in particular, we consider the Hamiltonian operator $\hat{H}(\xi)$,
which is dependent on the classical degrees of freedom $\xi$,
the hybrid dynamics is generated by the function 
$H(\eta,\xi) = \langle\psi(\eta)\vert \hat{H}(\xi)\vert \psi(\eta)\rangle$ and the Poisson bracket:
\begin{align}
\label{HamiltonHybridC}
\dot{\xi}_i &= \lbrace \xi_i, H \rbrace_H = \lbrace \xi_i, H \rbrace_C\,,
\\
\label{HamiltonHybridQ}
\dot{\eta}_a &= \lbrace \eta_a, H \rbrace_H = \lbrace \eta_a, H \rbrace_Q\,,
\end{align}
where in the last equality of both lines the classical and quantum nature of
$\xi$ and $\eta$ respectively has been invoked to make use of 
$\lbrace\xi,f\rbrace_Q = \lbrace\eta,f\rbrace_C =0 \;\;\forall f\in C^\infty(\mathcal{M})$.}

{\color{black}
One may further expand those equations: for the hybrid Poisson bracket acting on the classical variables $\xi_i$, one has:
\begin{equation}
\dot{\xi}_i =\lbrace \xi_i, H \rbrace_C =\sum_j
\left( (\partial_{Q_j}\xi_i)\langle \psi \vert\partial_{P_j}\hat{H} (\xi)\vert\psi\rangle 
- (\partial_{P_j}\xi_i)\langle \psi \vert\partial_{Q_j}\hat{H} (\xi)\vert\psi\rangle\right)\,.
\end{equation}
If one then considers the cases $\xi_i=Q_k$ or $\xi_i=P_k$ separately,
one arrives to Newton's-like equations for $Q_i$, $P_i$:
\begin{align}
\label{eq:ehrenfest1}
\dot{Q}_k &= \langle \psi \vert \frac{\partial \hat{H}}{\partial P_k}\vert\psi\rangle\,,
\\
\label{eq:ehrenfest2}
\dot{P}_k &= -\langle \psi \vert \frac{\partial \hat{H}}{\partial Q_k}\vert\psi\rangle\,,
\end{align}
}

{\color{black}
On the other hand, the dynamical equation of the the quantum variables,
\begin{equation}
\dot{\eta}_a =\lbrace \eta_a,H(\xi,\eta) \rbrace_Q\,,
\end{equation}
is exactly the  same as Eq.~(\ref{QuantumHamiltonEquation}), but with a dependence of the Hamiltonian operator
on the classical variables. Therefore, this equation, as it was shown in the previous section for the quantum-only case,
is equivalent to Schrödinger's equation for $\psi(\eta)$ -- although one must maintain that parametric dependence of the Hamiltonian
operator on the classical variables $\xi = Q, P$:
\begin{align}
\label{eq:ehrenfest3}
\dfrac{d}{dt}\vert{\psi} \rangle&= -i \hat{H}(Q, P)\vert\psi\rangle\,.
\end{align}}

{\color{black} Eqs.~(\ref{eq:ehrenfest1}-\ref{eq:ehrenfest2}) and (\ref{eq:ehrenfest3}) 
are Ehrenfest's equations for a hybrid model.} From our previous analysis
of the classical and the quantum case,  it becomes clear
that they compose a Hamiltonian system with the Poisson bracket
defined as in Eq.~(\ref{eq:hybpoisson}). Perhaps the form given in Eqs.~(\ref{eq:ehrenfest1}),
(\ref{eq:ehrenfest2}), and (\ref{eq:ehrenfest3}) is still not the most recognizable; if one
uses the Hamiltonian defined in Eqs.~(\ref{eq:enham1}) and (\ref{eq:enham2}) for a set
of electrons and nuclei, one gets:
\begin{align}
\label{eq:ehrenfest1bis}
\dot{\vec{Q}}_I &= \frac{\vec{P}_I}{M_I}
\\
\label{eq:ehrenfest2bis}
\dot{\vec{P}}_I &= -\langle \psi \vert \vec{\nabla}_I \hat{H}_{\rm e}(Q) \vert\psi\rangle\,,
\\
\label{eq:ehrenfest3bis}
\dfrac{d}{dt}\vert{\psi} \rangle&= -i \hat{H}(Q, P)\vert\psi\rangle\,.
\end{align}

Allured by the Hamiltonian character of this set of equations, one may be tempted to consider
the Gibbs equilibrium ensemble, Eq.~(\ref{eq:gibbs}), to be the hybrid canonical one. In terms of
the quantum-classical variables, it reads:
\begin{align}
\label{eq:hybridgibbs}
\rho_{\rm CC}(\eta,\xi) &= \frac{1}{Z_{\rm CC }(\beta)}e^{-\beta H(\eta,\xi)}\,,
\\
Z_{\rm CC}(\beta) &= \int{\rm d}\mu(\xi){\rm d}\mu(\eta) \;e^{-\beta H(\eta,\xi)}\,.
\end{align}
It is also a stationary ensemble in the hybrid case. The averages over this ensemble
would be the ones obtained if one attaches a thermostat tuned to temperature $T = \frac{1}{k_B \beta}$ to Ehrenfest dynamics, 
propagates a trajectory $(\eta^\beta(t),\xi^\beta(t))$, and
computes the time averages:
\begin{align}
\nonumber
\langle A \rangle_{\rm CC}(\beta) &= \lim_{t_f\to\infty}\frac{1}{t_f}\int_0^{t_f}\!\!{\rm d}t\; A(\eta^\beta(t),\xi^\beta(t))
\\
\label{eq:hybridergodic}
 &= \frac{1}{Z_{\rm CC }(\beta)}
\int{\rm d}\mu(\xi){\rm d}\mu(\eta) \;A(\eta,\xi) e^{-\beta H(\eta,\xi)}\,.
\end{align}
For example, one practical way to proceed is to use Langevin's
dynamics (although there are various other thermostat definitions that
have been invented over the years), that essentially consists in
substituting the equation for the force (\ref{eq:ehrenfest2bis}) by:
\begin{equation}
\dot{\vec{P}}_I = -\langle \psi \vert \vec{\nabla}_I \hat{H}_{\rm e}(Q) \vert\psi\rangle
- \beta \gamma \frac{\vec{P}_I}{M_I} + \vec{\eta}_I(t)\,,
\end{equation}
where $\vec{\eta}_I(t)$ are stochastic Gaussian processes that must verify:
\begin{align}
\langle \vec{\eta}_I(t)\rangle &= 0\,,
\\
\langle \eta_{I\alpha}(t) \eta_{J\beta}(t')\rangle &= 2\gamma\delta_{IJ}\delta_{\alpha\beta}\delta(t,t')\,.
\end{align}
The $\alpha,\beta$ indices run over the three spatial dimensions;
see for example Ref.~\citenum{Tuckerman2010} for details.
In any case, the values thus obtained are \emph{not} the ensemble average values that one would wish
to obtain, given above in Eq.~(\ref{eq:hybavg}),
hence the previously documented numerical failure of this approach -- 
see for example Refs.~\citenum{Mauri1993,Parandekar2005,Parandekar2006}.
It should be noted, however, that this fact by itself should not be
considered a failure of Ehrenfest dynamics -- inasmuch as the same
fact noted above for the quantum case cannot be considered a failure
of Schr{\"{o}}dinger equation. It results, once again, of the erroneous application
of a technique invented for purely classical systems to hybrid ones, that contain
some quantum variables.

The underlying reason behind the difference of the two ensembles is that, in classical systems,
all points in the phase space are mutually exclusive, and are given a Boltzmann weight in the canonical ensemble.
The MD procedure (and in particular the thermostats) was designed to produce a phase space visitation consistent with this.
In the hybrid phase space, due to the quantum character of one of its parts, not all distinct points are
mutually exclusive events,\cite{Alonso2020} and therefore the ensemble targetted by the thermostats, determined by a Boltzmann weight over the phase space, does not match the HC ensemble.


\subsection{\label{subsec:correctedehrenfest} Corrected averages for Ehrenfest dynamics}

Nevertheless, Eq.~(\ref{eq:hybridergodic}) can be useful, as we will
show now. The thermostatted Ehrenfest dynamics does sample the phase
space, and it generates an ensemble, even if wrong. One may
then apply a reweighting procedure -- essentially, modifying
the averaging in the time integral -- and obtain the
correct hybrid ensemble averages. This can be done in fact in several ways.

The first thing to notice is that Eq.~(\ref{eq:hybridergodic}) holds
for any function $g(\eta, \xi)$ on $\mathcal{M}$, not only on the ones
that result of a hybrid observable as $A(\eta, \xi) =
\langle\eta\vert\hat{A}(\xi)\vert\eta\rangle$. Then one may ask the
question: for any hybrid observable $\hat{A}$, can one find a function
$g_{\hat{A}}(\eta,\xi)$, such that:
\begin{equation}
\label{eq:gdef}
\langle\hat{A}\rangle_{\rm HC}(\beta) = \langle g_{\hat{A}}\rangle_{\rm CC}(\beta)\,?
\end{equation}
If so, one could then perform the dynamics and use Eq.~(\ref{eq:hybridergodic})
with $g_{\hat{A}}$ in order to obtain $\langle
g_{\hat{A}}\rangle_{\rm CC}(\beta)$, and therefore the \emph{true}
hybrid ensemble average $\langle\hat{A}\rangle_{\rm HC}(\beta)$.

The answer is positive, and there is not only one, but many possible
functions that can be used. In the following, we consider two examples:

\begin{enumerate}

\item Equation (\ref{eq:gdef}) holds if $g_{\hat{A}}$ is defined as:
\begin{align}
g_{\hat{A}}(\eta, \xi) &= \mu(\beta) e^{\beta H(\eta,\xi)}{\rm Tr}\left[ e^{-\beta\hat{H}(\xi)}\hat{A}(\xi)\right]\,,
\;\;\;\textrm{where}
\\
\mu(\beta) &= \frac{Z_{\rm CC}(\beta)}{(\int{\rm d}\mu(\eta))Z_{\rm HC}(\beta)}\,.
\end{align}
The computation of the normalization factor $\mu(\beta)$ may seem
problematic, but it can be obtained from the dynamical trajectory, in
the following way: For each $g_{\hat{A}}$, we define an ``unnormalized'' function
\begin{equation}
\tilde{g}_{\hat{A}}(\eta,\xi) = \frac{g_{\hat{A}}(\eta, \xi)}{\mu(\beta)} = 
e^{\beta H(\eta,\xi)}{\rm Tr}\left[ e^{-\beta\hat{H}(\xi)}\hat{A}(\xi)\right]\,,
\end{equation}
such that $\langle g_{\hat{A}}\rangle_{\rm CC}(\beta) = \mu(\beta) \langle \tilde{g}_{\hat{A}}\rangle_{\rm CC}(\beta)$.

On the other hand, we know that for the identity operator,
$\langle \hat{I}\rangle_{\rm HC}(\beta) = 1$, and therefore:
\begin{equation}
\langle g_{\hat{I}}\rangle_{\rm CC}(\beta) = 
\mu(\beta) \langle \tilde{g}_{\hat{I}}\rangle_{\rm CC}(\beta) = \langle \hat{I}\rangle_{\rm HC}(\beta) = 1\,.
\end{equation}
Thus, we may compute $\mu(\beta)$ as $1/\langle \tilde{g}_{\hat{I}}\rangle_{\rm CC}(\beta)$,
and $\langle \tilde{g}_{\hat{I}}\rangle_{\rm CC}(\beta)$ can be obtained from a dynamics propagation, i.e.:
\begin{equation}
\frac{1}{\mu(\beta)} = 
\lim_{t_f\to\infty}\frac{1}{t_f}\int_0^{t_f}\!\!{\rm d}t\; g_{\hat{I}}(\eta^\beta(t),\xi^\beta(t))\,.
\end{equation}
Summarizing, a final formula that permits to compute the hybrid ensemble averages is:
\begin{equation}
\label{eq:formula1}
\langle \hat{A}\rangle_{\rm HC}(\beta) = \lim_{t_f\to\infty}
\frac{
\int_0^{t_f}\!{\rm d}t\;
e^{\beta H(\eta^\beta(t),\xi^\beta(t))}{\rm Tr}\left[ e^{-\beta\hat{H}(\xi^\beta(t))}\hat{A}(\xi^\beta(t))\right]
}
{
\int_0^{t_f}\!{\rm d}t\;
e^{\beta H(\eta^\beta(t),\xi^\beta(t))}{\rm Tr}\left[ e^{-\beta\hat{H}(\xi^\beta(t))}\right]
}
\end{equation}

Therefore, the procedure consists of performing a thermostatted
Ehrenfest dynamics, and computing the previous time integrals over the
obtained trajectory $(\eta^\beta(t),\xi^\beta(t))$. One obvious difficulty lies in
the computation of the traces over the quantum Hilbert space, whose
difficulty depends on the level of theory used to deal with the quantum
electronic problem.

\item Equation (\ref{eq:gdef}) also holds if $g_{\hat{A}}$ is defined as:
\begin{align}
g_{\hat{A}}(\eta, \xi) &= \lambda(\beta) \sum_\alpha \delta(\eta - \eta_\alpha(\xi)) A_{\alpha\alpha}(\xi)\,,
\\
\lambda(\beta) &= \frac{Z_{\rm CC}(\beta)}{Z_{\rm HC}(\beta)}\,,
\end{align}
where $\eta_{\alpha}(\xi)$ are the adiabatic states:
\begin{equation}
\hat{H}(\xi)\vert\eta_\alpha(\xi)\rangle = E_\alpha(\xi)\vert\eta_\alpha(\xi)\rangle\,,
\end{equation}
and
\begin{equation}
A_{\alpha\alpha}(\xi) = \langle\eta_\alpha(\xi)\vert\hat{A}(\xi)\vert\eta_\alpha(\xi)\rangle\,.
\end{equation}
The difficulty due to the computation of the $\lambda(\beta)$ factor
can be solved in a similar way to the method used in the previous
case, leading to the following final formula:
\begin{equation}
\label{eq:formula2}
\langle \hat{A}\rangle_{\rm HC}(\beta) = \lim_{t_f\to\infty}
\frac{
\int_0^{t_f}\!{\rm d}t\;
\delta(\eta^\beta(t)-\eta_\alpha(\xi^\beta(t)))A_{\alpha\alpha}(\xi^\beta(t))
}
{
\int_0^{t_f}\!{\rm d}t\;
\delta(\eta^\beta(t)-\eta_\alpha(\xi^\beta(t)))
}\,.
\end{equation}
This formula avoids the need to compute all the electronic excited
states, necessary for the traces present in
Eq.~(\ref{eq:formula1}). In exchange, it contains a probably worse
numerical difficulty: the presence of the delta functions. The
interpretation of these is the following: during the trajectories, one
should not count in the average the state that is being visited,
unless the trajectory passes by an eigenstate of the Hamiltonian (a
state of the adiabatic basis). In other words, apart from the
normalization factor given by the denominator, this formula is a
modification of the straigthforward average given in
Eq.~(\ref{eq:hybridergodic}), that discards all states except for the
adiabatic eigenstates.

That correction is easy to understand intuitively. Let us first rewrite
the hybrid canonical ensemble density matrix,
\begin{equation}
\label{eq:rhohcemat}
\hat{\rho}_{\rm HC}(\xi) = \frac{1}{Z_{\rm HC}(\beta)}e^{-\beta \hat{H}(\xi)}\,,
\end{equation}
in terms of its spectral decomposition for each $\xi$:
\begin{equation}
\hat{\rho}_{\rm HC}(\xi)=\frac{1}{Z_{HC}(\beta)}\sum_{\alpha}e^{-\beta E_\alpha(\xi)} \hat \eta_\alpha(\xi),
\end{equation}
where $E_\alpha(\xi)$ are the eigenvalues, and $\hat \eta_\alpha(\xi)$ the projectors on the 
eigenspaces of the Hamiltonian $\hat H(\xi)$ (we assume, for simplicity, that there is no 
degeneration; otherwise one would just need to use an orthogonal basis for each degenerate subspace). 
Now, this expression can be written in terms of a (generalized) probability distribution function in $\mathcal{M}$, as:
\begin{equation}
\label{eq:rhohcepdf}
\rho_{\rm HC}(\eta,\xi) = \frac{1}{Z_{\rm HC}(\beta)} \sum_{\alpha}\delta(\eta - \eta_\alpha(\xi))e^{-\beta E_\alpha(\xi)}\,.
\end{equation}
This distribution determines  $\hat{\rho}_{\rm HC}(\xi)$, since:
\begin{equation}
\hat{\rho}_{\rm HC}(\xi) = \int{\rm d}\mu(\eta) \rho_{\rm HC}(\eta,\xi) 
\frac{\vert\eta\rangle\langle\eta\vert}{\langle\eta\vert\eta\rangle}\,.
\end{equation}
By comparing Eq.~(\ref{eq:rhohcepdf}) with Eq.~(\ref{eq:hybridgibbs}), it becomes clear that the error
that this latter equation does is counting all possible states, whereas the true hybrid ensemble only
counts the states in the adiabatic basis. Of course, one could choose a different basis, but the point
is that the ``classical'' Gibbs distribution (\ref{eq:hybridgibbs}) overcounts the quantum states. For
a deeper discussion on this issue, we refer the reader to Ref.~\citenum{Alonso2020}.

In order to implement this procedure numerically, one should of course
use some finite representation of the delta functions, giving them a
non-zero width. It is unclear, however, that this would lead to an
efficient scheme, since the propagation would probably have to be very
long in order to obtain an accurate sampling of the quantum states.

\end{enumerate}

\section{\label{sec:md} Approaches that do not require the propagation of the electrons}

The use of Ehrenfest dynamics to compute the ensemble averages, as
described in the previous section, has a notable caveat: it requires
the explicit propagation of the electrons. The time scale associated
to the electronic movement is very small (of the order of
attoseconds), which makes hybrid MD schemes computationally intensive
due to the need of a very fine time step.

In this section, we show how this problem can be circumvented by 
making use of dynamics that do not explicitly propagate the electrons,
such as ground-state Born-Oppenheimer MD -- including the necessary
correction to account for the hot electrons --, or the dynamics based
on the electronic free energy surface that has already been used in
the past. In this way, we frame these approaches into the
theoretical setup described above. 

Let us suppose that we perform a MD for the classical particles, based
on a Hamiltonian function $\mathcal{H}(\xi)$ (to be specified
below). In this case, the dynamics is not hybrid: the propagation
equations involve only the classical particles, moving under the
influence of $\mathcal{H}(\xi)$.  The ergodic assumption, if it holds,
permits to compute:
\begin{equation}
\langle g \rangle_{\rm CC}(\beta) = \frac{1}{Z_{\rm CC}(\beta)}\int{\rm d}\mu(\xi)\;e^{-\beta \mathcal{H}(\xi)}g(\xi)
 = \lim_{t_f\to\infty}\frac{1}{t_f}\int_0^{t_f}\!\!{\rm d}t\; g(\xi^\beta(t))\,,
\end{equation}
for any function $g(\xi)$. Notice that the classical canonical
ensemble that we are using now refers to the classical degrees of
freedom $\xi$ only, as opposed to the one used in the previous
section, that included the quantum ones.

In the same manner as we did in the
previous section, one may wonder the following: for a given hybrid observable
$\hat{A}(\xi)$, does there exist some function $g_{\hat{A}}(\xi)$ such that
\begin{equation}
\label{eq:gdef2}
\langle\hat{A}\rangle_{\rm HC}(\beta) = \langle g_{\hat{A}}\rangle_{\rm CC}(\beta)\,?
\end{equation}

Once again, the answer is affirmative, and in more ways than one. One obvious possibility,
analogous to the first one used for Ehrenfest dynamics, is:
\begin{align}
g_{\hat{A}}(\xi) &= \mu(\beta) e^{\beta\mathcal{H}(\xi)} {\rm Tr}\left[e^{-\beta\hat{H}(\xi)}\hat{A}(\xi)\right]\,,
\\
\mu(\beta) &= \frac{Z_{\rm{CC}}(\beta)}{Z_{\rm{ HC}}(\beta)}\,.
\end{align}
As it happened in the previous section, the calculation of the
normalization factor $\mu(\beta)$ does not require of the explicit
computation of the partition functions (that may be impractical),
but may result from the MD propagation itself, using the identity
$\langle\hat{I}\rangle_{\rm HC}(\beta) = 1$. Using this fact and the same
procedure shown in the previous section, one arrives to the final formula:
\begin{equation}
\label{eq:formula3}
\langle \hat{A}\rangle_{\rm HC}(\beta) = \lim_{t_f\to\infty}
\frac{
\int_0^{t_f}\!{\rm d}t\;e^{\beta\mathcal{H}(\xi^\beta(t))}{\rm Tr}\left[e^{-\beta\hat{H}(\xi^\beta(t))}\hat{A}(\xi^\beta(t))\right]
}{
\int_0^{t_f}\!{\rm d}t\;e^{\beta\mathcal{H}(\xi^\beta(t))}{\rm Tr}\left[e^{-\beta\hat{H}(\xi^\beta(t))}\right]
}\,.
\end{equation}
This formula is very similar to Eq.~(\ref{eq:formula1}). However, the
trajectory $\xi^\beta(t)$ to be used here must be obtained through a
thermostatted classical-only MD determined by a Hamiltonian function
$\mathcal{H}(\xi)$, in contrast to the hybrid quantum-classical
Ehrenfest dynamics used in the previous section.  The Hamiltonian
function $\mathcal{H}(\xi)$ is in fact arbitrary, although a bad
choice for this object could lead to a very bad convergence with
respect to the total propagation time $t_f$ -- since using the ergodic 
hypothesis requires an accurate sampling of phase space. Two options that
immediately come to mind are:

\begin{enumerate}

\item Using the electronic free energy:
\begin{equation}
\mathcal{H}(\xi) = F(\xi;\beta) = -\frac{1}{\beta}\log{\rm Tr}e^{-\beta\hat{H}(\xi)}\,.
\end{equation}
This actually permits to simplify Eq.~(\ref{eq:formula3}) into a very appealing form:
\begin{equation}
\label{eq:formula4}
\langle \hat{A}\rangle_{\rm HC}(\beta) = \lim_{t_f\to\infty} \frac{1}{t_f}
\int_0^{t_f}\!{\rm d}t\; \langle \hat{A}(\xi^\beta(t))\rangle_{\rm Q}\,,
\end{equation}
where at each classical phase space point in the trajectory one must compute the quantum ensemble average:
\begin{equation}
\langle \hat{A}(\xi)\rangle_{\rm Q} = 
\frac{{\rm Tr}\left[\hat{A}(\xi)e^{-\beta\hat{H}(\xi)}\right]}
{{\rm Tr}e^{-\beta\hat{H}(\xi)}}
\end{equation}
Eq.~(\ref{eq:formula4}) reminds of the usual MD ergodic averaging formula, just substituting the
observable value by the thermal quantum average. In fact, if the observable that one is interested
in is purely classical, $\hat{A}(\xi) = A(\xi)\hat{I}$, the formula is identical:
\begin{equation}
\label{eq:formula5}
\langle A \rangle_{\rm HC}(\beta) = \lim_{t_f\to\infty} \frac{1}{t_f}
\int_0^{t_f}\!{\rm d}t\; A(\xi^\beta(t))\,.
\end{equation}
Therefore, for purely classical observables, if one uses the
electronic free energy instead of the ground-state adiabatic energy as
the Hamiltonian driving the ionic movement, the resulting MD provides
the hybrid canonical averages using the ``standard'' ergodic
average. If the observable is itself hybrid, one must compute at each
point during the trajectory the quantum thermal average.

This propagation of the classical variables following the electronic
free energy surface underlies the scheme put forward by Alavi
\latin{et al.}~\cite{Alavi1994}, although in that work the procedure
is tightly tied to the use of FTDFT as the scheme that handles the
electronic structure problem (computation of the free energy, and of
its gradients). The same concept was also suggested by
some of the current authors in Refs.~\citenum{Alonso2010,Alonso2012}.

\item Using the ground-state Born-Oppenheimer energy:
\begin{equation}
\mathcal{H}(\xi) = E_0(\xi)
\end{equation}
In this case we would just need to do the usual ground-state
Born-Oppenheimer MD, which has the advantage of being a very well
known and tested technique, for which plenty of codes and tools
exist. In order to obtain the hybrid ensemble averages that do not
ignore the electronic temperature, however, one must use the averaging
formula (\ref{eq:formula3}), which for this case can be transformed
into:
\begin{equation}
\label{eq:formula6}
\langle \hat{A}\rangle_{\rm HC}(\beta) = \lim_{t_f\to\infty}
\frac{
\int_0^{t_f}\!{\rm d}t\;\sum_\alpha e^{-\beta\Omega_\alpha(\xi^\beta(t))}A_{\alpha\alpha}(\xi^\beta(t))
}{
\int_0^{t_f}\!{\rm d}t\;\sum_\alpha e^{-\beta\Omega_\alpha(\xi^\beta(t))}
}\,.
\end{equation}
where $\alpha$ runs over all the adiabatic eigenstates, and 
\begin{equation}
\Omega_\alpha(\xi) = E_\alpha(\xi)-E_0(\xi)
\end{equation}
are the electronic excitations.

On top of the usual ground-state Born-Oppenheimer MD, the added
difficulty here would be the computation of these excitations, which
may be more or less demanding depending on the level
of theory used to model the many-electron problem. 

Note that if the observable $\hat{A}(\xi)$ is actually a classical
observable $A(\xi)\hat{I}$, this scheme can also be rewritten as:
\begin{equation}
\label{eq:formula6class}
\langle A\rangle_{\rm HC}(\beta) = \lim_{t_f\to\infty}
\frac{
\int_0^{t_f}\!{\rm d}t\; A(\xi^{\beta}(t)) e^{-\beta F(\xi^{\beta}(t);\beta)-E_0(\xi^{\beta}(t))}
}{
\int_0^{t_f}\!{\rm d}t\; e^{-\beta F(\xi^{\beta}(t);\beta)-E_0(\xi^{\beta}(t))} 
}\,.
\end{equation}
Here, we also write the formula in terms of the free energy. Computationally, the
difference with respect to the previous approach given in formula
(\ref{eq:formula5}) is that one does not need the gradients of
the free energy, necessary in the previous approach for the
computation of the forces in the dynamics.

\end{enumerate}

All previous formulas have assumed that the thermostat is fixed to the
target temperature. However, the dynamics can be performed at
a different temperature (a technique that has been used in MD to probe
larger regions of configuration space in less simulation time), as
long as the reweighting corrects for this. Take, for example, formula
(\ref{eq:formula3}), that we repeat here for convenience, although we
now use two different temperatures $\beta$ and $\beta'$:
\begin{equation}
\label{eq:formula7}
\langle \hat{A}\rangle_{\rm HC}(\beta) = \lim_{t_f\to\infty}
\frac{
\int_0^{t_f}\!{\rm d}t\;e^{\beta'\mathcal{H}(\xi^{\beta'}(t))}
{\rm Tr}\left[e^{-\beta\hat{H}(\xi^{\beta'}(t))}\hat{A}(\xi^{\beta'}(t))\right]
}{
\int_0^{t_f}\!{\rm d}t\;e^{\beta'\mathcal{H}(\xi^{\beta'}(t))}
{\rm Tr}\left[e^{-\beta\hat{H}(\xi^{\beta'}(t))}\right]
}\,.
\end{equation}
In this formula the temperature dependence is twofold:
\begin{enumerate}
    \item The temperature used to define the thermostat, which appears in the $\beta'$ labeling
the trajectories $\xi^{\beta'}(t)$. This should be equal to the temperature used for defining
the re-weighting factors, i.e. the inverse Boltzmann weight $e^{\beta' \mathcal{H}(\xi^{\beta'}(t))}$.
    \item The ``target'' temperature, that is the one that should be used in the  exponent 
of the un-normalized hybrid canonical ensemble density matrix, appearing inside the trace, $e^{-\beta\hat H}$.
\end{enumerate}
These two temperatures can be different, and formula
(\ref{eq:formula7}) still holds. In practical applications, this would
permit to obtain hybrid canonical ensemble averages at different
temperatures $\beta$ by computing a single thermostatted trajectory at
a fixed ``ergodic temperature'' $\beta'$.  This can also be done
when using Ehrenfest dynamics, and formulas (\ref{eq:formula1}) and
(\ref{eq:formula2}) above.  It does not hold, however, if one uses
formulas (\ref{eq:formula4}) or (\ref{eq:formula5}) for the dynamics
on the free energy surface, since they rely on a cancellation that is only
achieved if the two temperatures are equal (two temperatures can also
be used when doing the dynamics on the free energy, but one would then
need to compute the free energy at those two temperatures).  Note also
that, in practice, this procedure cannot be indefinitely extended to
any temperature range, since the ergodic visitation will not be effective 
unless the two temperatures are similar.

\mbox{}

{\color{black}

Summarizing, the previous formulas permit to use well known MD
techniques and obtain canonical averages that correctly
account for the electronic temperature. Looking, for example,
at Eq.~(\ref{eq:formula6class}), the procedure entails two steps:
\begin{enumerate}
    \item One first performs a standard first principles MD simulation using, for example, the common
technique based on ground-state DFT.
    \item Then, either on the fly as the trajectory is being generated, or later in a post-processing 
procedure, one computes the electronic free-energy at the trajectory points, using the finite-temperature DFT extension.
With such information, one can use Eq.~(\ref{eq:formula6class}) to correct the time averages that,
without this averaging method, would fail to converge to the real canonical ensemble.
\end{enumerate}
This scheme can be applied on top of trajectories obtained previously,
had they been saved.  Note that, for the reasons explained above, one
may recycle trajectories obtained with ground-state BOMD at some
(nuclear only) temperature, to compute ensemble averages at various
different global temperatures. This may be an advantage over the
procedure implied by Eq.~(\ref{eq:formula5}) (MD with forces computed
on the electronic free-energy surface), as in that case one trajectory
must be generated at each temperature. Finally, of course DFT need not be
the method to be used for the computation of the forces -- and the
finite temperature DFT need not be the procedure to obtain the free
energy, as one may use for example TDDFT to compute the electronic excitations
and apply formula (\ref{eq:formula6}).
}

\mbox{}
\subsection{Numerical example}

\begin{figure}
\begin{center}
  \includegraphics[width=0.5\linewidth]{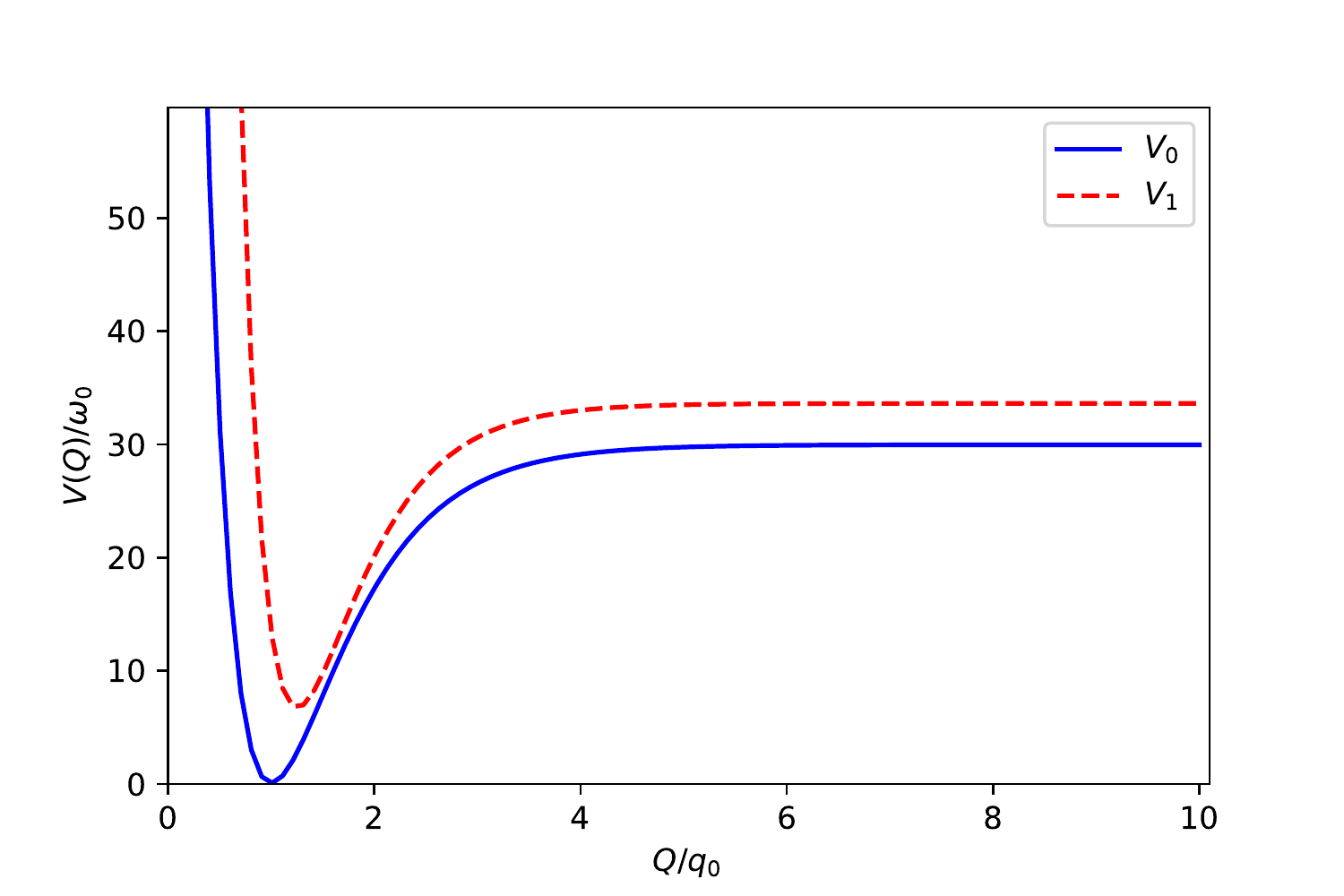}
  \caption{Electronic potential energy curves of the modeled dimer (in units of the ground state vibrational
frequency $\omega_0$): ground state (blue) and first excited state (dashed red), 
as a function of the dimer length (in units of the ground state equilibrium distance).}
  \label{fig:MorsePotential}
\end{center}
\end{figure}

We finish with a numerical demonstration of the validity of the formulas given above, using a simple
model and the very last of the presented schemes: ``standard''
ground-state Born-Oppheheimer MD with a correction formula. Thus, 
we consider a simple dimer model, using the internuclear distance $Q$ as the 
only classical position variable (being $P$ the corresponding momentum),
and the subspace generated by the two lowest electronic adiabatic states 
as the quantum space. Futhermore, we consider that these
two adiabatic states correspond to Morse potentials. Hence, the Hamiltonian operator 
ruling the hybrid dynamics can be written, in the basis of its eigenstates, as:
\begin{equation}
\label{Hamiltonian}
\hat H(Q, P)=\dfrac{P^2}{2m}\hat{\mathbb{I}}+\begin{pmatrix}
V_0(Q) & 0\\
0& V_1(Q)
\end{pmatrix}
\end{equation}
where $m$ is the dimer reduced mass, and the Morse potentials are given by:
\begin{equation}
V_i(Q) = D_i(1-e^{-b_i(Q-q_i)})^2+\Delta_i\,.
\end{equation}
The parameters defining the Morse potential for the $i$-th energy
level $V_i(Q)$ have an easy interpretation: $\Delta_i$ is a global
shift that sets the value of the curve at its minimum; $q_i$ the
position at that minimum ($V_i(Q=q_i)=\Delta_i$); $D_i$ defines how
quickly the potential ascends for $Q>q_i$, and also determines the
value of the gap between the minimum ($\Delta_i$) and the big $Q$
limit of the potential:
$\lim_{Q\rightarrow\infty}V_i(Q)=\Delta_i+D_i$. Lastly, $b_i$ defines
how narrow the well is, how sharply it grows when $Q\rightarrow 0$,
and also how rapidly it reaches the plateau for $Q>q_i$.
The vibrational frequency associated to each potential well
is given by $\omega_i = \sqrt{\frac{2b^2D}{m}}$.
Fig.~\ref{fig:MorsePotential} depicts these potential energy
curves~\bibnote{We supply, as supporting information, a computational
notebook containing all the code that generates the results displayed
in the article. It also contains all the chosen parameter values.}.


Using ground-state Born-Oppenheimer MD 
means that the classical degrees of freedom follow the Hamiltonian system defined by the function:
\begin{equation}
\mathcal{H}(Q, P) = E_0(Q, P) = \frac{P^2}{2m} + V_0(Q)\,.
\end{equation}
The system is coupled to a Langevin thermostat, at the temperature
given by $T=\frac{1}{k_B\beta}$. This dynamics provides an ergodic
curve over the classical phase space with a visitation weight given by
the Boltzmann factor $e^{-\beta E_0(Q, P)}$. Using the corrected
averaging procedure defined by Eq.~(\ref{eq:formula6}), we can obtain
the hybrid canonical ensemble averages. Although this formula is valid
for any observable, we chose to compute the average value of the
length of the dimer, a purely classical observable: $\hat{A}(Q, P) =
Q\hat{I}$.


\begin{figure}[h]
\begin{center}
  \includegraphics[width=0.75\linewidth]{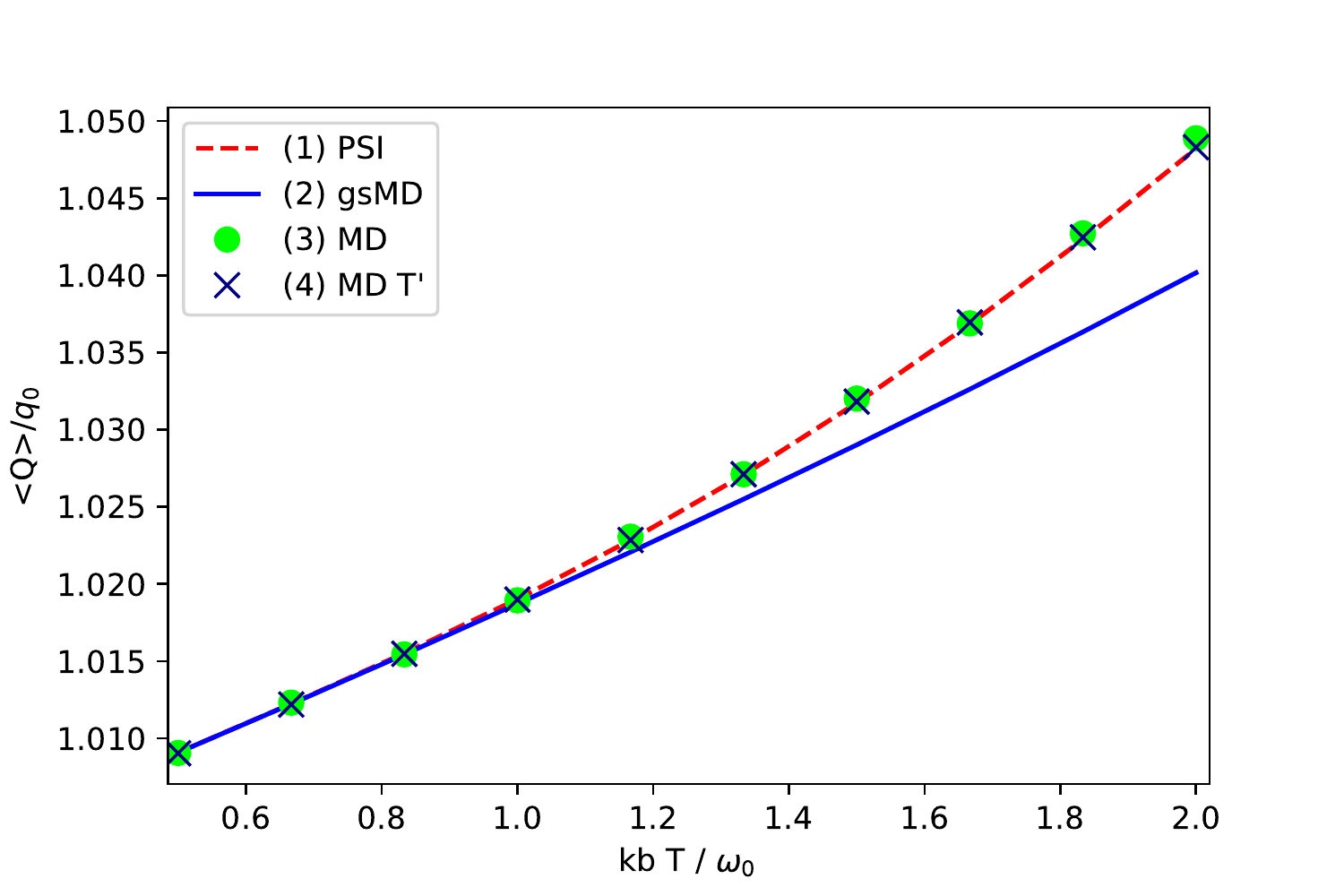}
  \caption{\label{fig:CorrectionFactors}
Ensemble average dimer length, $\langle Q\rangle_{\rm HC}(\beta)$, in units of the
ground state equilibrium length $q_0$, as a function of temperature, calculated via:
(1) direct integration in the phase space, labeled PSI;
(2) ground-state MD without
the correction, i.e. ignoring the electronic temperature, labeled gsMD; (3) ground-state MD with the application of the correcting average formula (\ref{eq:formula6}), 
labeled MD; (4) ground-state MD at a single temperature for the thermostat with the application of the correcting average formula (\ref{eq:formula7}) for the whole range of temperatures of the target HCE.
}
\end{center}
\end{figure}

Fig.~\ref{fig:CorrectionFactors} shows the results. As the model is
particularly simple, we can display both the exact values (i.e. the
hybrid canonical ensemble averages computed by performing the direct
integration in phase space, using Eq.~(\ref{eq:hybavg})), and the
values produced by using the time-averages over the dynamics. In this
latter case, we display both the corrected averages, that result from
formula (\ref{eq:formula6}), and the ergodic averages using
ground-state molecular dynamics, which corresponds to the
purely-classical canonical ensemble average.
It is clear how the proposed reweighting formula yields
the correct numbers.

We stress that, in the procedure presented above, the computation of
the ergodic trajectory is an indirect way to perform phase space
integrals over the classical phase space. In principle, any infinite
($t_f\rightarrow\infty$) ergodic trajectory could be used as a basis
to apply the corrected averaging procedure, as long as the implicit
distribution over the phase space that results of the dynamics is
compensated in the time averages: the use of the ground-state
potential energy surface to generate the dynamics is one of the
possible many choices. This only holds if the trajectory provides a
dense enough visitation of the phase space.

However, in practice, the simulations provide only finite-time
trajectories and, therefore, the visitation of phase space is not
dense. The effectiveness of a given dynamics will depend on what
regions of phase space it probes more frequently. Some thermostatted
MD trajectories will be more cost-effective than others, if they visit
more frequently the regions of the phase space that are
relevant to the target distribution.

In our example, the target distribution is the hybrid canonical
ensemble, and one should choose a dynamics that is likely to force the
system to spend time on its high probability regions. This is not the
only factor to consider, however. For example, it is likely that
Ehrenfest dynamics fulfills this condition, but the cost of
propagating Ehrenfest equations is high, due to the need to propagate the electrons. Likewise, it may happen that using the free energy as
the driving Hamiltonian is costly due to the requirement of computing
its gradients with respect to the classical degrees of freedom in order to obtain the forces. A dynamics that requires
longer times $t_f$ to achieve the convergence of the time average can
be however computationally cheaper if the cost of performing the
propagation itself is lower. We consider that ground-state MD can be a good compromise, 
specially at low temperatures, but the analysis strongly depends on the particular model, 
the electronic structure method, etc.

\section{\label{sec:conclusions} Conclusions}

We have examined the problem of computing the canonical ensemble
averages through MD calculations, for hybrid quantum classical systems
(typically, quantum electrons and classical nuclei in molecular or
condensed matter physics and chemistry). If the temperature is high
enough so that the electronic excited states cannot be ignored,
performing ground state Born Oppenheimer MD and computing the ergodic
averages on the generated trajectories does not yield the correct
ensemble averages.

The fact that one cannot assume that the electrons are inert,
adiabatically adapting to the ground state, naturally seems to demand
for a truly hybrid dynamics, such as Ehrenfest's. However, the
addition of a thermostat to these equations, followed by the
computation of the resulting observable time-averages, does not
produce the right averages either. The quantum character of part of
the electrons cannot be handled by the standard MD + thermostat
procedure, designed to produce essentially classical equilibrium
ensembles.

Nevertheless, performing a thermostatted dynamics, be it Ehrenfest, or
a purely classical one such as ground state MD, does generate a
trajectory (i.e. an ensemble) in phase space that can be reweighted in
order to obtain the true hybrid averages. This amounts to correcting
the time averaging formulas. It has been the purpose of this work to
examine here the various options that exist, setting them in a common
language. The procedure can be followed using Ehrenfest dynamics,
which is a hybrid dynamics, but can also be followed using
classical-only dynamics driven by, for example, the ground-state
Born-Oppenheimer Hamiltonian. Likewise, this framework does include,
as a particular case, the possibility of performing the nuclear
dynamics on the Hamiltonian that results of considering the electronic
free energy. In this case, if one is interested in computing averages
of purely classical observables, the time averages do not require
correction, as the factors cancel out. The suitability of any of these
procedures over the others depends on the particular model and level
of theory used to handle the electronic structure problem.



\begin{acknowledgement}



The authors acknowledge
financial support by MINECO Grant FIS2017-82426-P.
C. B.  acknowledges financial support by Gobierno de
Aragón through the grant defined in ORDEN IIU/1408/2018. 

\end{acknowledgement}
{\color{black}
\begin{suppinfo}
\begin{itemize}
\item md.nbconvert.pdf: The code used in the numerical example is provided and commented in this PDF exported from a Python notebook. One can find there the propagation of the dynamics described in the example, coupled with Langevin thermostat, as well as the time averaging procedures, both with the correction proposed in this paper and in the usual application of the ergodic principle. Allowed by the simplicity of the system, the HCE expected value of observables are also computed directly on phase space. 
\end{itemize}
\end{suppinfo}
}




\bibliography{manuscript}

\end{document}